\documentclass[%
 reprint,
 amsmath,amssymb,
 aps,
 prd
]{revtex4-2}
\usepackage{graphicx}
\usepackage{hyperref}
\usepackage{dcolumn}
\usepackage{bm}
\usepackage{epstopdf}
\usepackage{mathrsfs}
\usepackage{graphicx}
\usepackage{multirow}
\usepackage{amsmath}
\usepackage{threeparttable}
\usepackage{xcolor}
\usepackage{float}
\begin{document}

\preprint{arXiv:0000.00000}

\title{Mirage Sources and Large TeV Halo-Pulsar Offsets: Exploring the Parameter Space}

\author{Yiwei Bao$^{1,2}$}
\author{Ruo-Yu Liu$^{1,3}$}\email{ryliu@nju.edu.cn}
\author{Gwenael Giacinti$^{2,4}$}\email{gwenael.giacinti@sjtu.edu.cn}
\author{Hai-Ming Zhang$^{1}$}
\author{Yang Chen$^{1,3}$}\email{ygchen@nju.edu.cn}
\affiliation{$^1$Department of Astronomy, Nanjing University, 163 Xianlin Avenue, Nanjing 210023, China\\
$^2$Tsung-Dao Lee Institute, Shanghai Jiao Tong University, Shanghai 201210, China\\
$^3$Key Laboratory of Modern Astronomy and Astrophysics, Nanjing University, Ministry of Education, Nanjing, China\\
$^4$School of Physics and Astronomy, Shanghai Jiao Tong University, Shanghai 200240, China
}

\date{\today}

\begin{abstract}
We investigate the asymmetric propagation of 100\,TeV electrons (whose radiation mainly concentrates on 20--30\,TeV) in turbulent magnetic fields around pulsars, using GPU-accelerated simulations to explore their trajectories and interactions within pulsar wind nebulae and the interstellar medium. Key results include the identification of ``mirage'' sources indicating significant offsets in high-energy emissions from their originating pulsars, challenging the results of traditional symmetric diffusion models. By varying parameters like source distance, magnetic field strength, and electron injection spectral index, the study delineates their effects on observable phenomena such as the identified number of sources, as well as the source separation. Our results offer insights into some puzzling sources observed recently by the Large High Altitude Air Shower Observatory (LHAASO), and shed light on the cosmic-ray transport mechanism in the interstellar medium.
\end{abstract}
\maketitle
\section{Introduction}
Pulsars are rapidly rotating neutron stars. They power relativistic pulsar winds composed of $e^\pm$ pairs \cite{2014ApJ...783L..21S,2020A&A...642A.204C,2021Sci...373..425L} at the expense of their rotational energies. A termination shock is formed where the pulsar wind encounters the ambient medium and reaches a pressure balance. Either at the termination shock \cite{2011ApJ...741...39S}, or far upstream of the termination shock \cite{2017A&A...607A.134C}, magnetic reconnection may occur and some of the $e^\pm$ pairs be accelerated up to $\sim$ PeV \cite{2021Sci...373..425L}. Alternatively, these particles may be accelerated at the termination shock, especially in its equatorial region~\cite{2018ApJ...863...18G,2020A&A...642A.123C}. The radiation of these relativistic pairs in the shocked wind gives rise to the pulsar wind nebula (PWN). 

Recent observations of extended TeV emissions around middle-aged pulsars by HAWC \cite{2017Sci...358..911A} and LHAASO \cite{2021PhRvL.126x1103A} have sparked significant interest among the community. It can be inferred from the surface brightness profiles of these extended $\gamma$-ray emission that $e^\pm$ pairs are escaping their PWNe and diffusing in the ambient interstellar medium (ISM) with a significantly suppressed diffusion coefficient, by a factor of $\sim 100-1000$, compared to the Galactic mean value. It may be interpreted as an enhancement of the interstellar turbulence \cite{Evoli18, FangKun21,Mukhopadhyay22}, or a reasonably large level of pre-existing turbulence~\cite{2018MNRAS.479.4526L}, or ascribed to the cross-field transport of particles if the local magnetic field and the observer's line of sight is approximately aligned \cite{2019PhRvL.123v1103L}. The size of the region with suppressed particle diffusion is suggested to be at least 20\,pc \cite{2023PhRvD.107l3020S}. The extended $\gamma$-ray emission region beyond the PWN is usually referred to as ``pulsar halos''. 

In principle, each middle-aged pulsar could power a TeV halo \cite{2017PhRvD..96j3016L, 2019PhRvD.100d3016S}. Indeed, we see many sources identified by LHAASO \cite{2024ApJS..271...25C} and HAWC \cite{2020ApJ...905...76A} are associated with middle-aged pulsars. However, many of them present significant spatial offsets between positions of the pulsars and the centroid of the TeV sources, which is not expected if pairs are injected from the position of the pulsar (or from the compact PWN around the pulsar) and isotropically diffuse in the surrounding ISM \citep{ZhangYi21}.
Actually, isotropic diffusion is a good approximation of the real transport of particles only if the physical scale considered in the problem is larger than multiple coherence lengths of the turbulent magnetic field \cite{2012PhRvL.108z1101G} (about 1/5 of the outer scale of the turbulence for Kolmogorov spectra), such as e.g. in the modelling of the global transport of cosmic rays (CRs) in the Galaxy. 

However, if we focus on the particle transport around accelerators, especially those located above/below the Galactic disk or in the outer Galaxy region, the isotropic diffusion approximation could be problematic, given that the average coherent length ($L_{\rm c} \approx 1/5$ of the injection length scale) of the ISM is of the order of $\sim 100$\,pc \cite{2016JCAP...05..056B}. This is particularly important for pulsar halo candidates detected by LHAASO-KM2A, because these sources are due to the emission from electrons/positrons with energies above 100\,TeV. Given the rapid cooling rate of these extremely energetic leptons, they may probably cool down before crossing one magnetic coherence length. Therefore, the distribution of high-energy particles could be highly asymmetric, following the direction of the local magnetic field. In the companion paper \citep{paperI} (Paper I), we have shown that the asymmetric propagation of electrons/positrons around their accelerators may cause multiple ``mirage'' sources and significant spatial offset between the accelerator and the source. The main goal of the present paper is to give a more quantitative description of our method, and to explore extensively the influence of different model parameters on the obtained source morphology. 

The rest of the paper is organized as follows: in Section II, we describe our method in detail; we explore the properties of the simulated sources with different parameters in Section III; In Section IV, we discuss several possible applications of our findings, and we conclude in Section V.



\section{Method Description}
To characterise the asymmetric propagation of particles, we conduct test-particle simulations by calculating the trajectories of 100\,TeV electrons, the radiation spectrum of which typically peaks at approximately {\bf 1}0\,TeV via the inverse Compton (IC) scattering off the cosmic microwave background (CMB). We employ a first-principle approach by injecting electrons into a turbulent magnetic field and determining their motion based on the Lorentz force \cite{1999ApJ...520..204G,2007JCAP...06..027D,2012JCAP...07..031G}. We employ GPU-acceleration methods, allowing the code to operate in a highly parallelized manner. Here, we thoroughly explain in detail the process of conducting the simulation in terms of particle motion and field generation.

\subsection{Generating Turbulent Magnetic Fields}

The propagation and energy loss of cosmic rays are critically influenced by the Galactic magnetic field, which comprises a large-scale regular component and a small-scale turbulent component. This work implements a sophisticated model for the turbulent magnetic field to simulate the stochastic deflections and scattering of charged particles.

\subsubsection{Field Generation Method}

The turbulent magnetic field \(\mathbf{B}_{\text{turb}}(\mathbf{x})\) is generated numerically using a Fourier-space method that ensures the field to be divergence-free (\(\nabla \cdot \mathbf{B} = 0\)), a fundamental property of physical magnetic fields. The model is characterized by the following steps:

\begin{enumerate}
    \item \textbf{Power Spectrum:} The magnetic energy density is distributed according to a power-law spectrum in wavenumber space, \(E(k) \propto k^{n_{\text{inx}}}\). Our simulation adopts a spectral index of \(n_{\text{inx}} = -11/3\), consistent with Kolmogorov-type turbulence. The spectrum is generated between a minimum wavenumber \(k_{\text{min}}\) and a maximum (cut-off) wavenumber \(k_{\text{cut}}\).

    \item \textbf{Polarization and Phase:} For each wavenumber \(\mathbf{k}\) in the 3D Fourier grid, the magnetic field component \(\mathbf{B}(\mathbf{k})\) is constructed using a polarization vector \(\mathbf{e}(\mathbf{k})\) that is orthogonal to \(\mathbf{k}\) (ensuring \(\mathbf{k} \cdot \mathbf{B}(\mathbf{k})=0\)). The field amplitude is scaled by the power spectrum, and a random phase is assigned to each mode. Gaussian random numbers are used to set the stochastic amplitudes.

    \item \textbf{Hermitian Symmetry:} To guarantee that the inverse Fourier transform yields a real-valued field \(\mathbf{B}_{\text{turb}}(\mathbf{x})\) in configuration space, the complex amplitudes in Fourier space are constrained by Hermitian symmetry, \(\mathbf{B}(-\mathbf{k}) = \mathbf{B}^*(\mathbf{k})\).

    \item \textbf{Inverse Fourier Transform:} The field in real space is obtained via a three-dimensional inverse Fast Fourier Transform (FFT), executed by a routine based on the Numerical Recipes algorithm. The resulting field is defined on a discrete, periodic lattice of \(256^3\) grid points.
\end{enumerate}

\subsubsection{Multi-Scale Turbulence Implementation}

To capture a realistic inertial range spanning several decades in scale, the total turbulent field is modeled as the superposition of four independent turbulent fields. Each component is generated with an identical Kolmogorov spectrum but at different spatial grid resolution (denoted as $\delta_1$, $\delta_2$, $\delta_3$, $\delta_4$), effectively sampling different scale ranges. By taking the grid spacings \(\delta_1 = 4\times10^{-9}\)~kpc, \(\delta_2 = 2\times10^{-7}\)~kpc, \(\delta_3 = 1\times10^{-5}\)~kpc, and \(\delta_4 = 5\times10^{-4}\)~kpc, this multi-grid approach allows for an effective dynamic range far exceeding that of a single \(256^3\) grid. The total turbulent field at any point in space is the sum of the fields interpolated from these four grids. The overall strength of the turbulent field is controlled by the parameter \(\sigma_{B,\text{uG}}\), which scales the root-mean-square (RMS) field strength. This comprehensive model enables a realistic simulation of cosmic ray trajectories and energy evolution in a turbulent magnetized medium.

\subsection{Simulating Particle Propagations}
We employ the original Boris pusher which avoids the phase error \citep[see e.g., Boris A pusher in][]{2018PhPl...25k2110Z} to preserve phase space volume. Since the electric field in the turbulent plasma may be neglected \cite{1999ApJ...520..204G}, the Boris pusher can be written as 
\begin{align}
  \boldsymbol{x}':&= \boldsymbol{x}_{n-1}+ \frac{1}{2} \boldsymbol{v}_{n-1} \cdot \Delta t \\
  \boldsymbol{t} :&= \tan\frac{\theta}{2}\hat{\boldsymbol{b}}\\
  \gamma\boldsymbol{v}':&= \gamma\boldsymbol{v}_{n-1} + \gamma\boldsymbol{v}_{n-1} \times \boldsymbol{t}\\
  \gamma\boldsymbol{v}_{n} &= \gamma\boldsymbol{v}' + \frac{2}{1+|\boldsymbol{t}|^2}\left(\gamma\boldsymbol{v}'\times \boldsymbol{t}\right)\\
  \boldsymbol{x}_n &= \boldsymbol{x}'+ \frac{1}{2} \boldsymbol{v}_{n} \cdot \Delta t 
\end{align}
where $\hat{\boldsymbol{b}}:= \boldsymbol{B}/|\boldsymbol{B}|$ is the unit vector, $\gamma$ the electron Lorenrz factor, $n$ the time step, $\boldsymbol{x}'$, $\boldsymbol{t}$, $\boldsymbol{v}'$ are three intermediate variables. A time step of 1/100 gyration period is utilized in the simulation. We also take synchrotron/IC losses into consideration. 
The synchrotron loss of electrons with Lorentz factor $\gamma$ reads
\begin{equation}
\dot{\gamma}_{\rm sync} = -3.25\times 10^{-8} \frac{B^2}{8\pi}\gamma^2,
\end{equation}
where $B$ is the local magnetic field strength. For IC losses we use the approximation given by \citet{2021ChPhL..38c9801F}:
\begin{equation}
\dot{\gamma}_{\rm IC} = -\frac{20c\sigma_{\rm T}\omega\gamma^2}{\pi^4m_ec^2}Y(\gamma,T),
\end{equation}
where $\omega$ is the grey-body energy density, $T$ the grey-body temperature and $Y(\gamma,T)$ the parameterised approximation function \cite{2021ChPhL..38c9801F}. For the IC seed photons, we use the interstellar radiation field averaged over a 2 kpc region surrounding the Sun as a reference, which is composed of five gray-body components with temperatures of 23209 K, 6150.4 K, 3249.3 K, 313.3 K, and 33.1 K \cite{2021ChPhL..38c9801F}. These components have corresponding energy densities of 0.12, 0.23, 0.37, 0.055, and 0.25 eV cm$^{-3}$ \cite{2021ChPhL..38c9801F}, respectively.

In our numerical scheme for simulating the evolution of an electron population, we implement a weighted macro-particle approach to efficiently sample the initial energy spectrum, which is depicted as follows:

First, the initial electron spectrum is discretized into \( N = 8192 \) bins, equally spaced in logarithmic energy between a minimum \( E_{\text{min}} \) and a maximum \( E_{\text{max}} \). Instead of injecting a large number of particles per bin, we inject a single \textit{macro-particle} at the origin (\( \mathbf{x} = 0 \)) at time \( t=0 \) for each bin. This approach significantly reduces computational cost while maintaining statistical fidelity.

Second, each macro-particle is assigned a statistical weight \( w_i \) proportional to the number of physical electrons in its respective energy bin. For an initial differential energy spectrum of the form
    \begin{equation}
        \frac{dN}{dE} = k E^{-\alpha} \exp(-E/E_{\text{cut}}),
    \end{equation}
    the number of electrons in a bin of width \( dE \) is \( (dN/dE) \, dE \). Using the relation \( d\log E = dE / E \), the number of electrons per logarithmic bin is
    \begin{equation}
        \frac{dN}{d\log E} = E \frac{dN}{dE} \propto E^{1-\alpha} \exp(-E/E_{\text{cut}}).
    \end{equation}
    Therefore, the weight for a particle representing a bin centered at energy \( E \) is assigned as
    \begin{equation}
        w_i \propto E^{1-\alpha} \exp(-E/E_{\text{cut}}).
    \end{equation}
    The weights are subsequently normalized such that their sum represents the total number of particles in the simulated population.

    Then, all macro-particles are simultaneously propagated through the combined regular and turbulent magnetic fields by numerically integrating the relativistic equation of motion. During propagation, their energies are continuously degraded due to synchrotron and inverse Compton losses, which are calculated at each integration step based on the local magnetic field strength and the particle's instantaneous energy.

    At last, the complete state of each macro-particle, including its position \( \mathbf{x} \), energy \( E \), and statistical weight \( w \), is recorded at fixed time intervals (e.g., every 10 years). Each recorded state is treated as an independent sample in post-processing analysis. This procedure effectively builds a large statistical ensemble from the initial 8192 particles, allowing for the detailed reconstruction of the spectral and spatial evolution of the entire electron population over time.

This combination of logarithmic energy binning, statistical weighting, and temporal sampling provides a robust and efficient framework for simulating the non-linear evolution of cosmic-ray electrons under the influence of transport and radiative cooling.

\section{Simulation data analysis}

\subsection{Methods}

\begin{figure*}[htbp]
    \centering
    \includegraphics[width=0.4\linewidth]{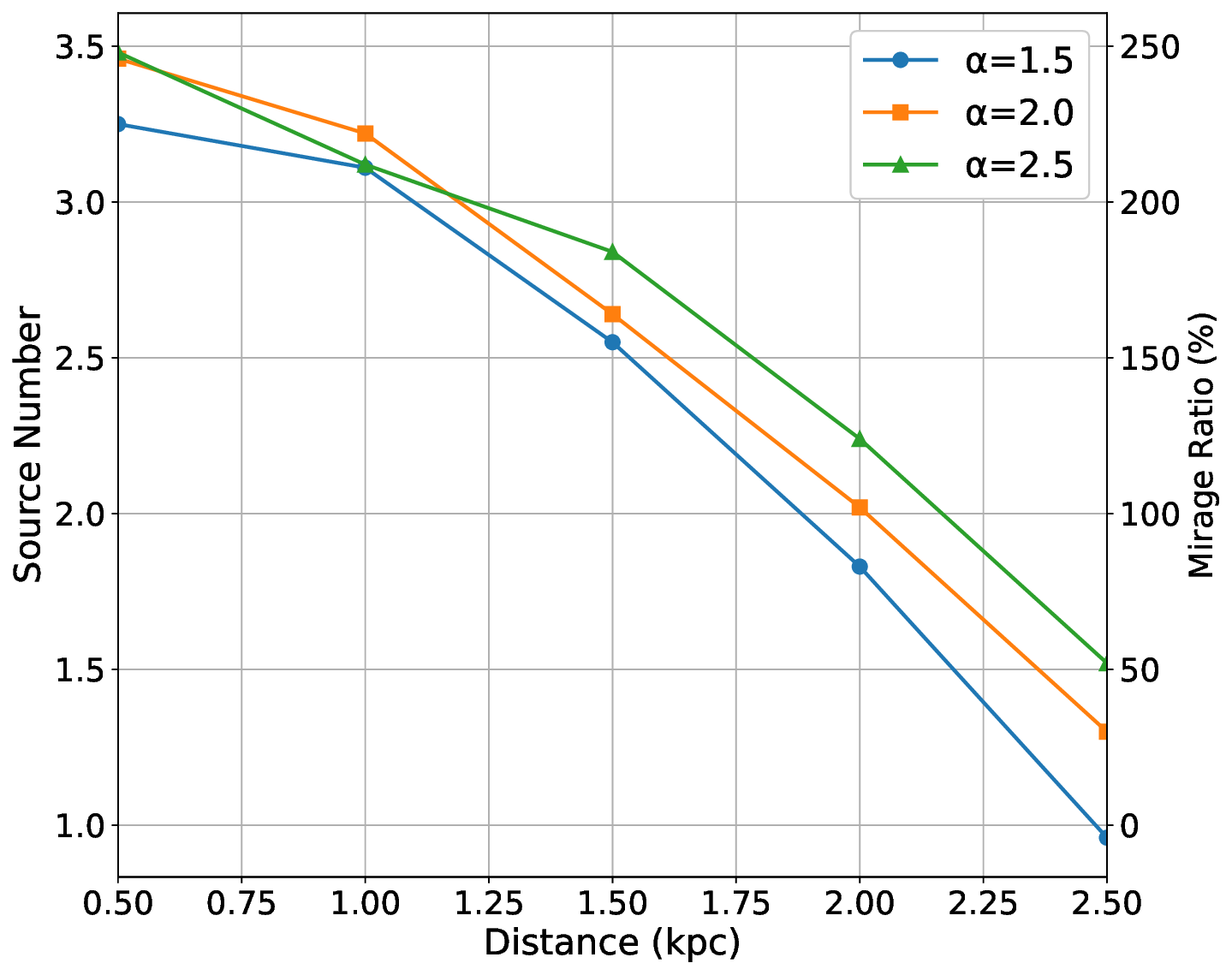} 
    \includegraphics[width=0.4\linewidth]{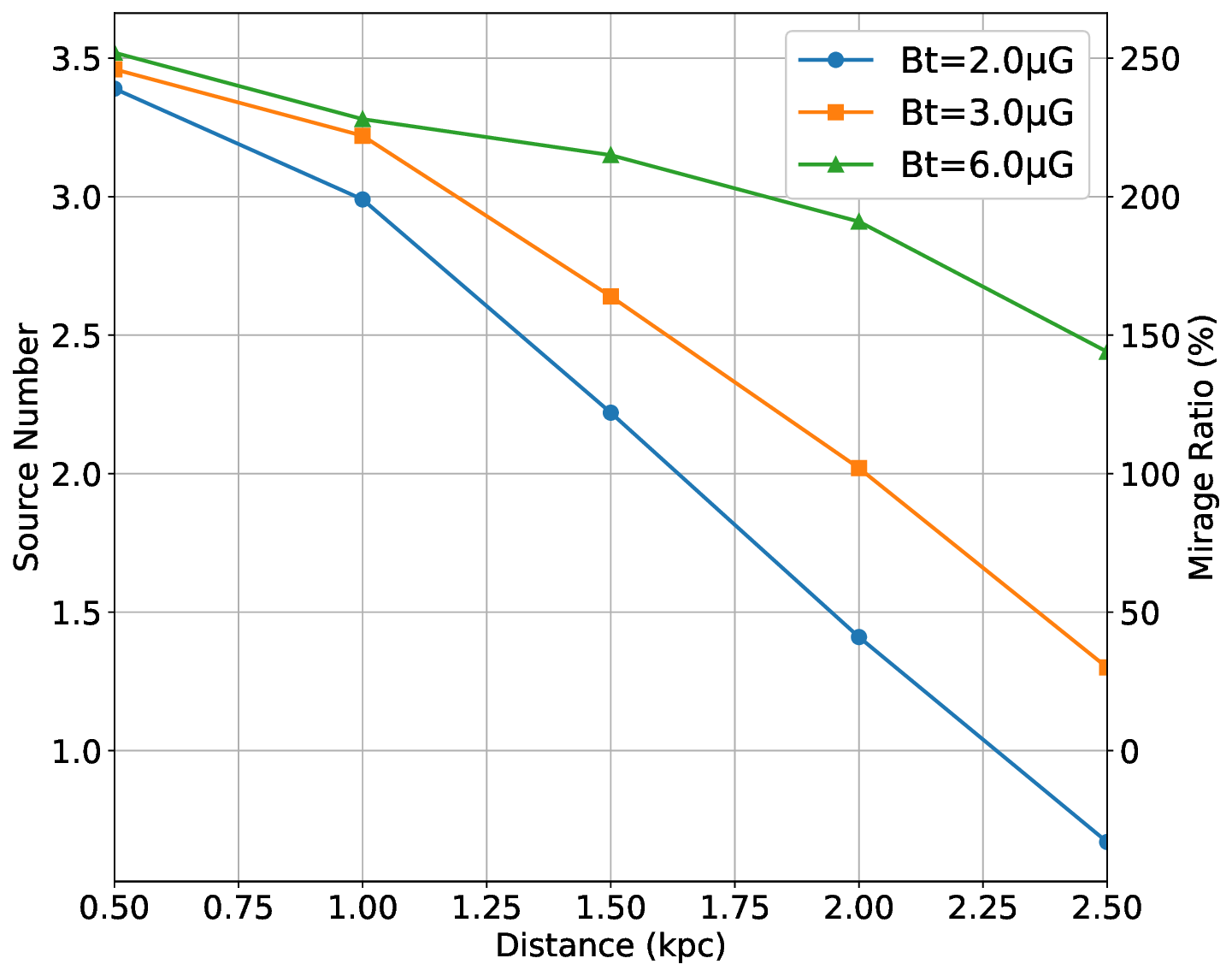}\\   
    \includegraphics[width=0.4\linewidth]{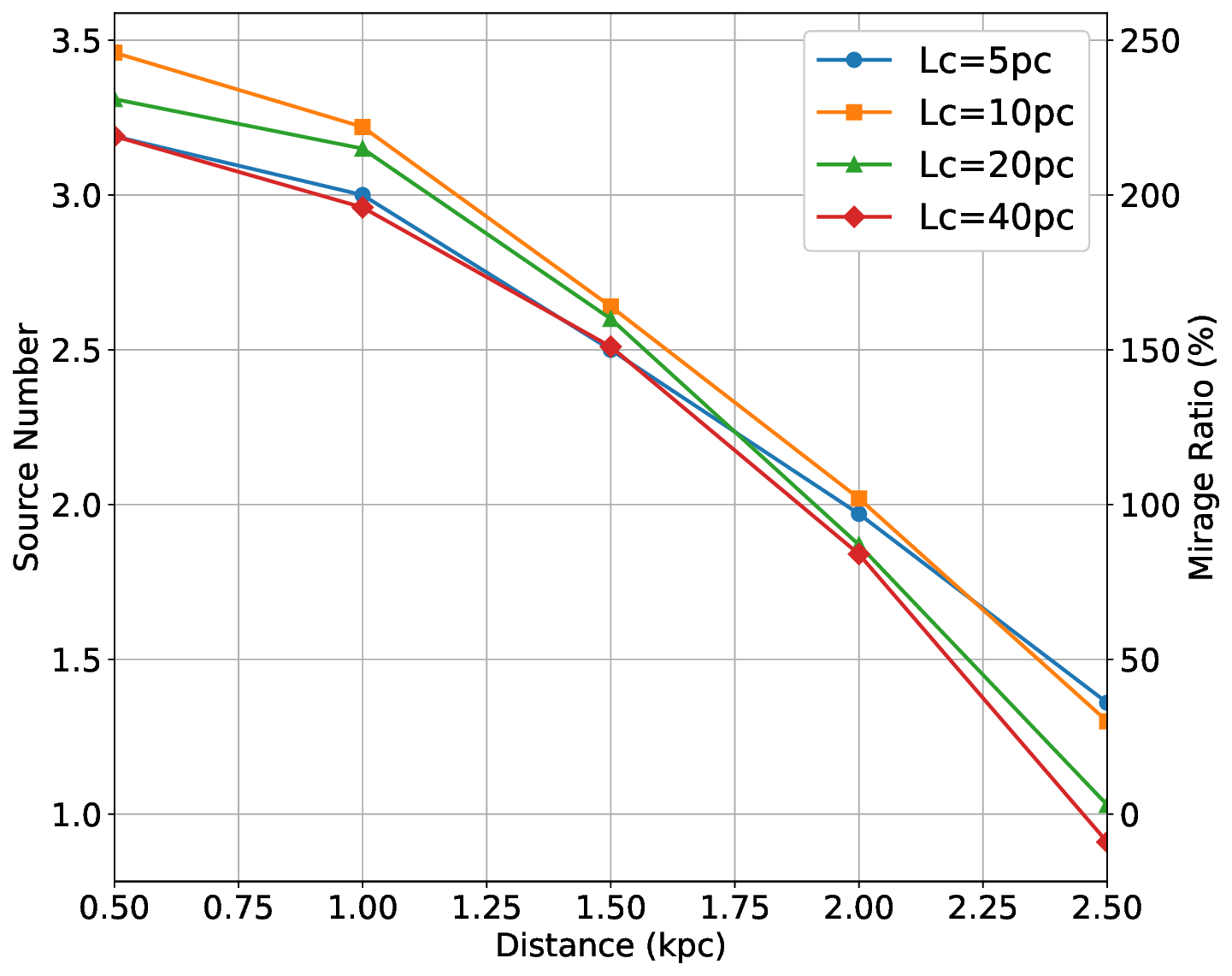}
    \includegraphics[width=0.4\linewidth]{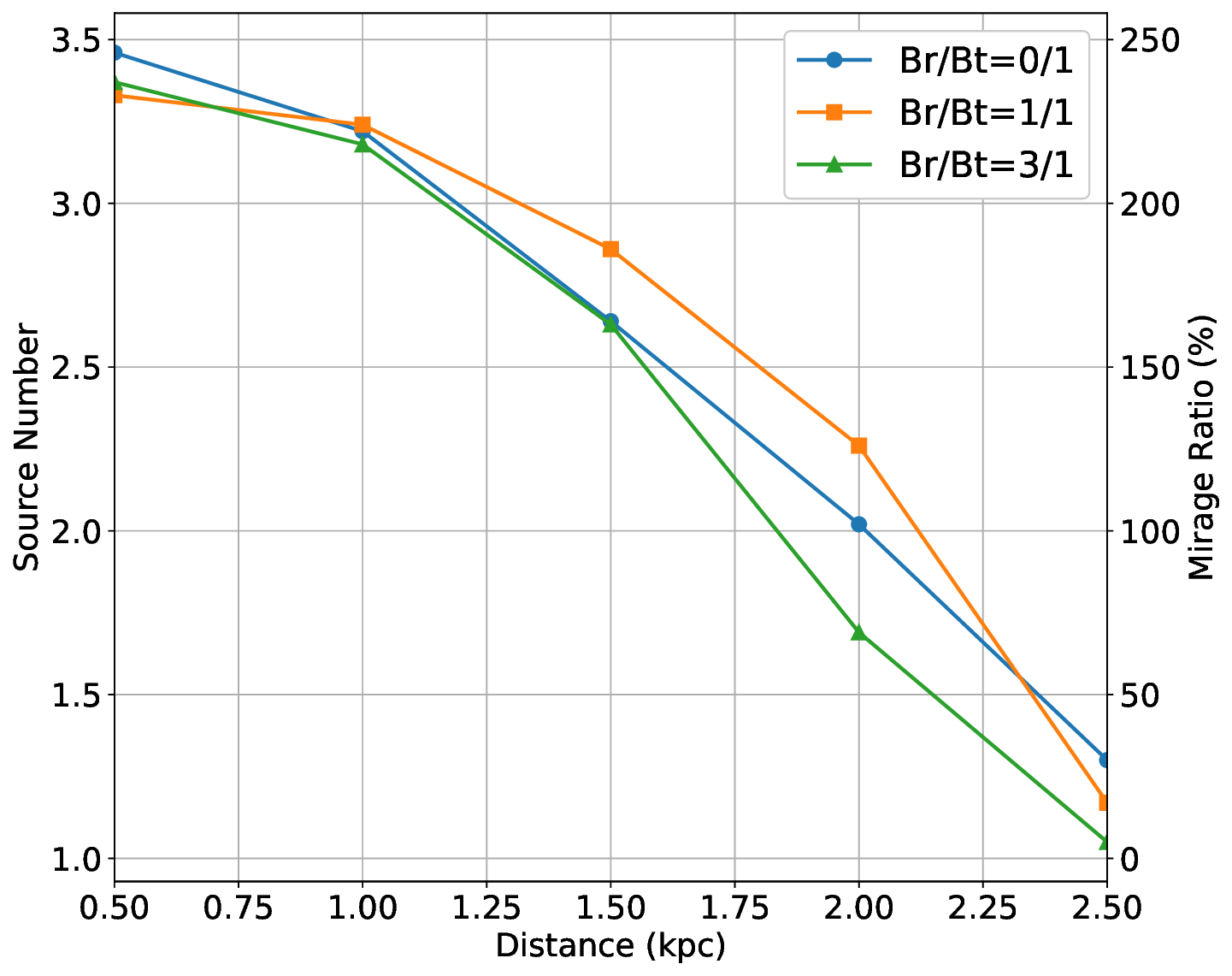}\\
    \caption{
        Variation of the source numbers $\xi$ with different plasma parameters: (a) electron injection index $\alpha$, (b) total strength of the magnetic field, (c) coherence length of the turbulent field and (d) regular-to-turbulent ratio. The results demonstrate how $\xi$ depends on different parameters and provide insights into the characteristic scale of turbulence structures in the plasma.\label{fig:xi}
    }
\end{figure*}

\begin{figure*}[htbp]
    \centering
    \includegraphics[width=0.4\linewidth]{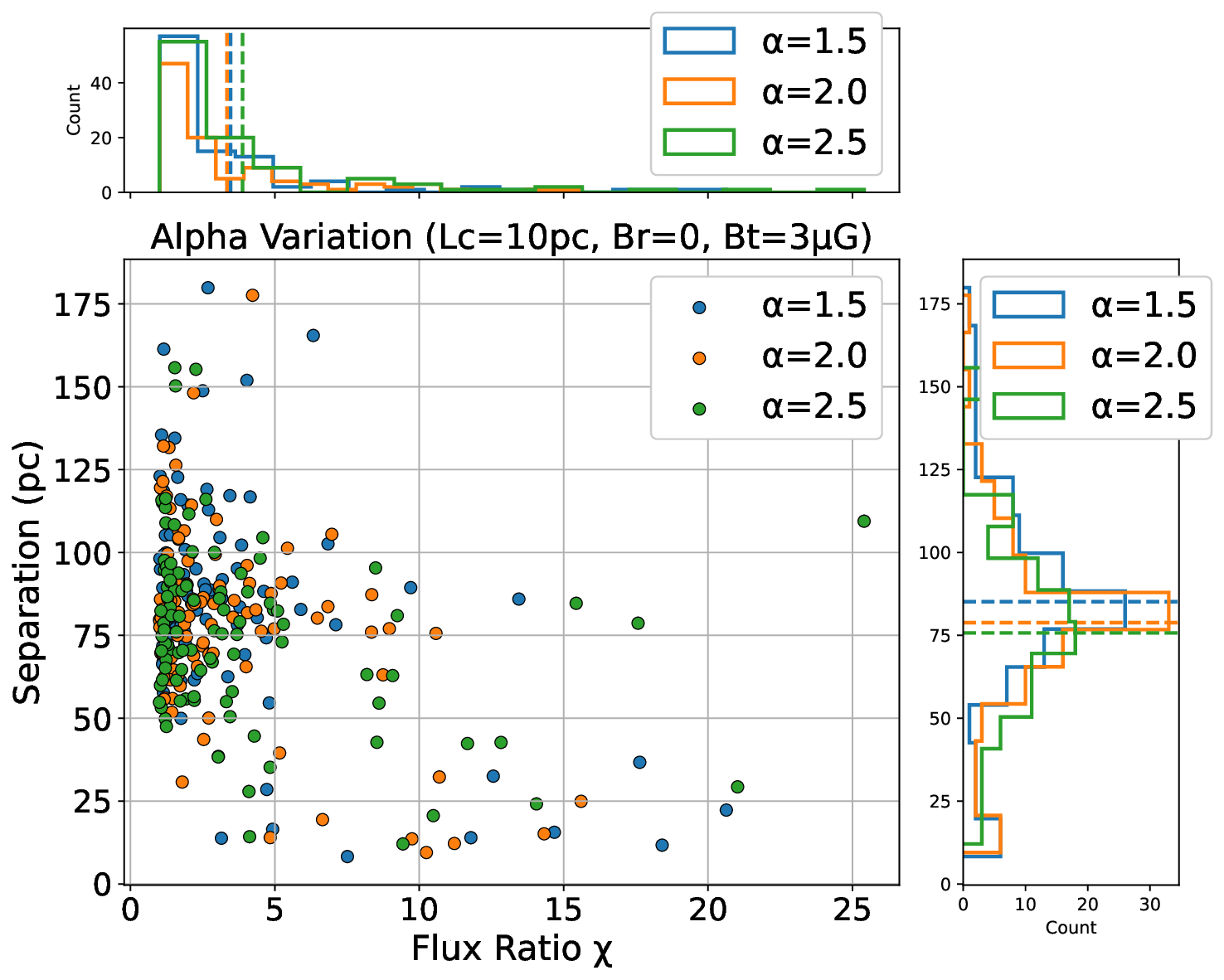}
    \includegraphics[width=0.4\linewidth]{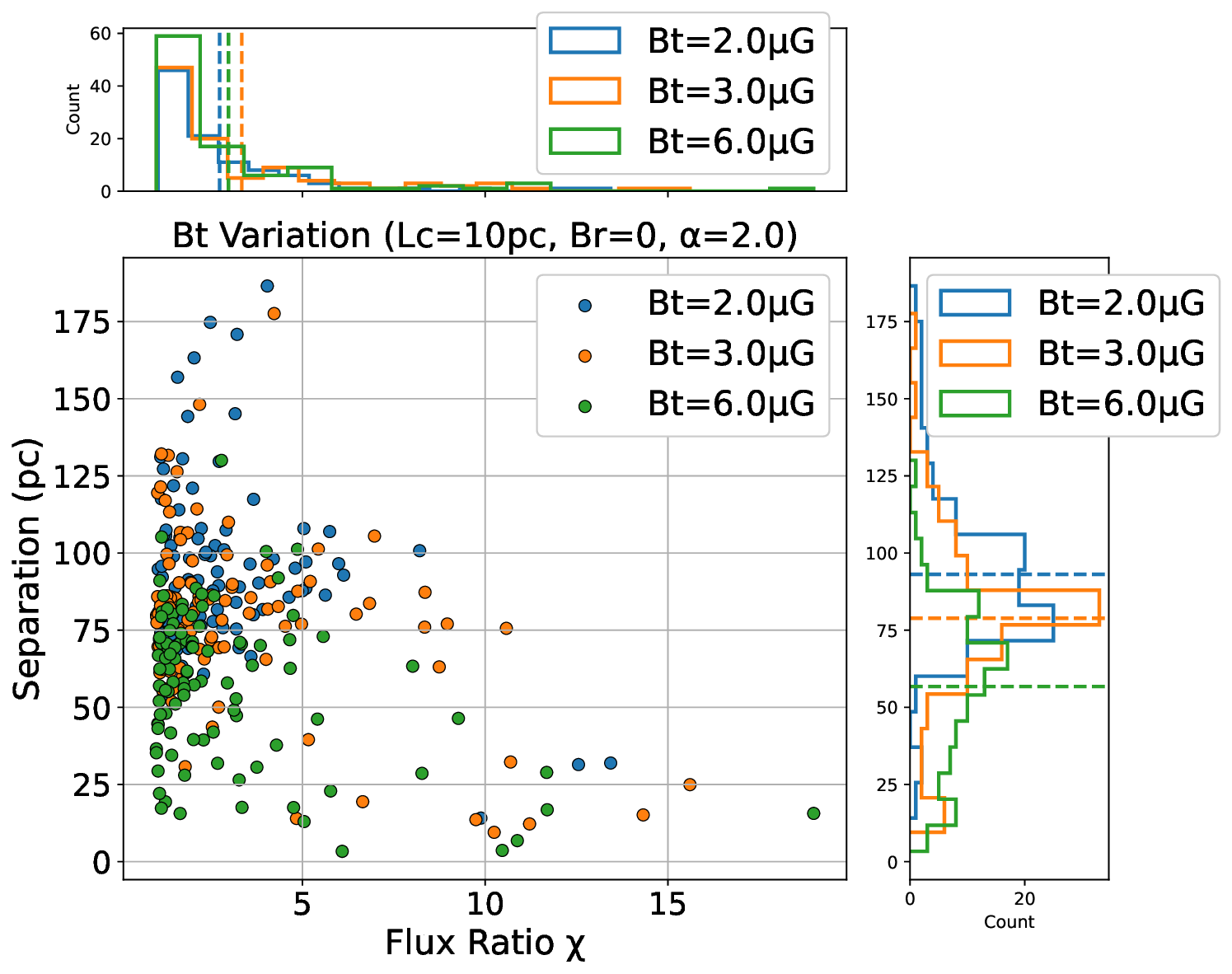}\\
    \includegraphics[width=0.4\linewidth]{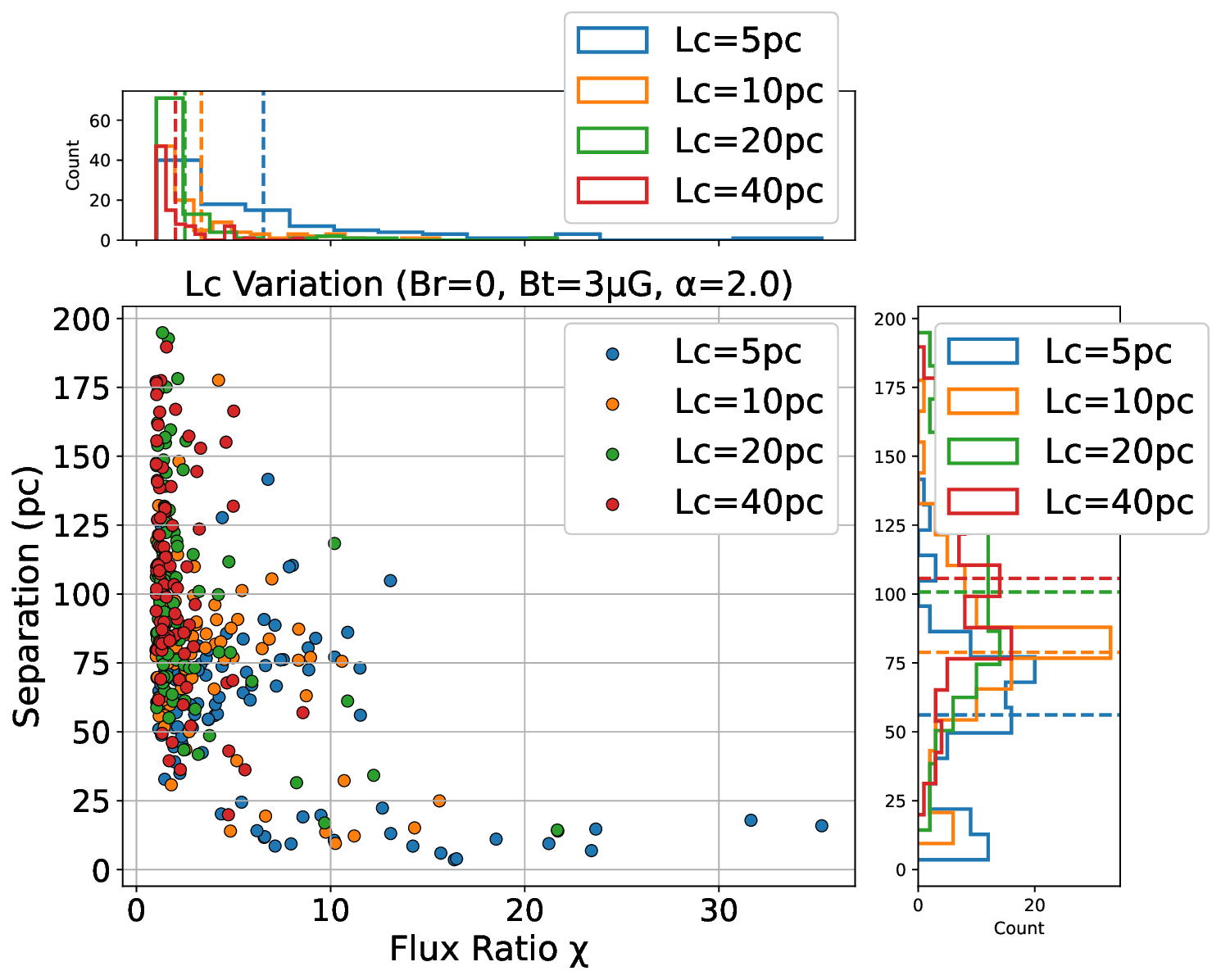} 
    \includegraphics[width=0.4\linewidth]{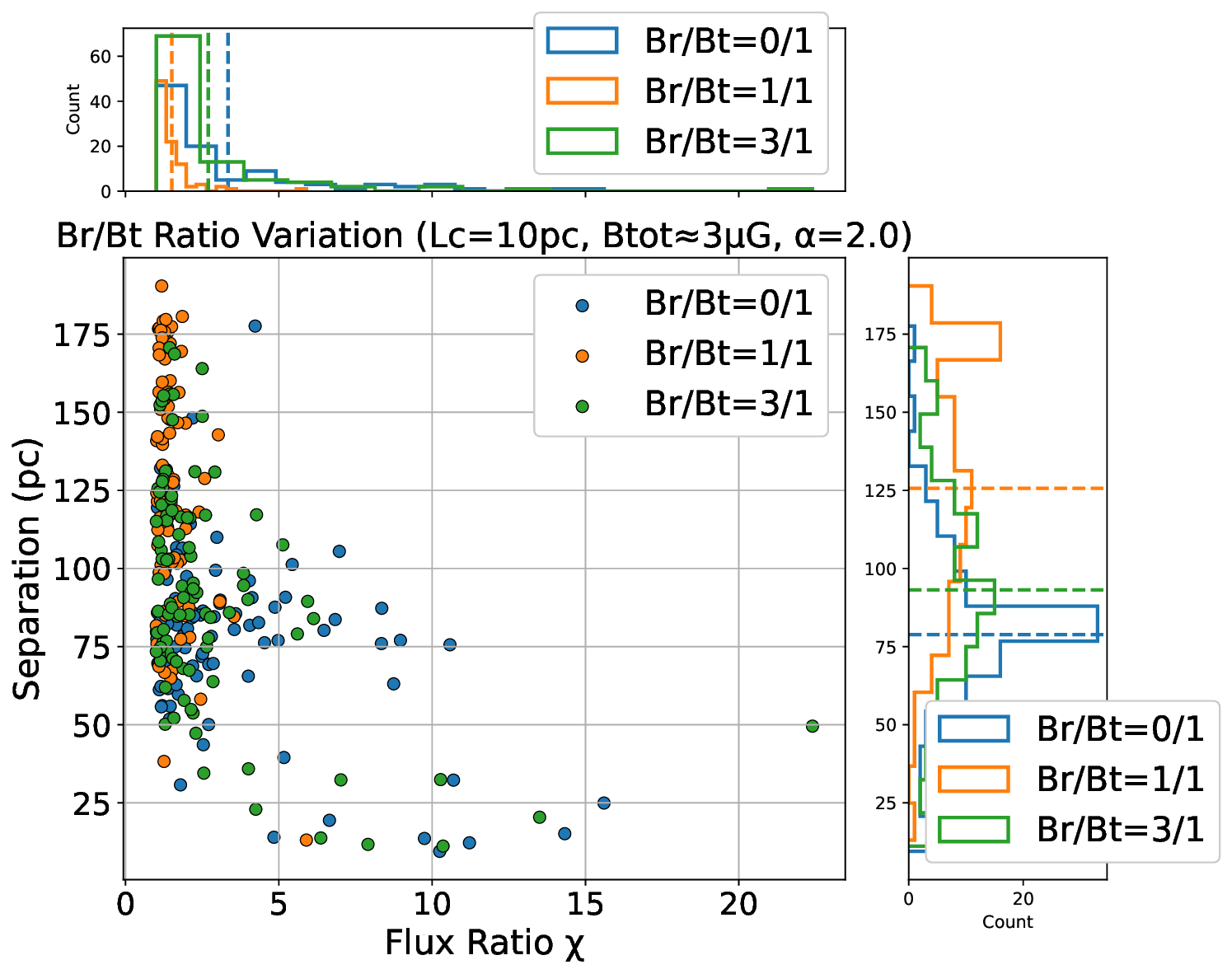} 
    \caption{Scatter plot of the flux ratio $\chi$ and separation distance for different $L_{\rm c}$ (top panel), magnetic field strength $B^2$ (middle panel) and the strength ratio between regular field and turbulent field $B_r/B_t$ (bottom panel). The distance is fixed at 1 kpc. The dashed line in the histogram mark the average value of the separation. \label{fig:Separation}}
\end{figure*}

To compare our simulation results with observational data, we generate synthetic gamma-ray source image as would be observed by LHAASO. The point spread function (PSF) of the instrument is modeled as a two-dimensional Gaussian with a standard deviation of $\sigma_{\rm PSF} = 0.43^\circ$, consistent with the typical angular resolution of current-generation LHAASO around 30\,TeV. For our analysis, we focus on the very-high-energy gamma-ray flux in the $25-40$\,TeV energy band, which provides an optimal balance between high signal statistics and superior angular resolution for morphological studies.

We normalize the total synthetic gamma-ray flux from the simulated electron population to a reference luminosity of $L_{\rm ref} = 6 \times 10^{32}$\,erg/s spanning over 5 orders of magnitude, corresponding to a total event rate of approximately {$1.6\times 10^6 (d/0.25\,{\rm kpc})^{-2}(A_{\rm eff}/1\,\rm km^2)$, assuming a power-law index of 2.0} counts per year with $d$ the distance of the pulsar and $A_{\rm eff}$ the effective area of the instrument. This normalization is scaled proportionally with $(250\,{\rm pc}/d)^2$ for sources at different distances. Our region of interest (ROI) encompasses a $400\,{\rm pc} \times 400\,{\rm pc}$ field centered on the injection site, divided into spatial pixels of $0.1^\circ \times 0.1^\circ$. At the reference distance of $d = 250$\,pc, this angular size corresponds to a physical scale of $0.436\,{\rm pc} \times 0.436\,{\rm pc}$ per pixel.

The simulated electron distributions are converted to gamma-ray counts through an Inverse Compton scattering matrix that accounts for interactions with the cosmic microwave background and interstellar radiation fields. The resulting photon counts in each spatial and energy bin are then convolved with the instrumental PSF and combined with a uniform background noise model. The background level is estimated from off-source regions and follows Poisson statistics, with an average rate of $\sim 0.318$ counts per hour per pixel at 0.1\,TeV, scaled appropriately for our observation time and energy range.

To quantify the source morphology in the synthetic maps, we employ a Bayesian model comparison framework using Markov Chain Monte Carlo (MCMC) sampling. We fit the observed count distribution with models containing an increasing number of Gaussian sources, selecting the optimal model through and a likelihood ratio test with a threshold of $\Delta \log \mathcal{L} > 12.5$ (TS $> 25$) for significant detection of additional sources. This approach allows us to objectively determine whether the simulated gamma-ray emission is best described by an extended halo or multiple distinct components.

To mimic the background, we generate a Poisson noise map (noise A) based on the background event rate measured from the Crab Nebula region, which is 1{\bf 0} event per hour within a 1$^\circ$ cone \cite{2021Sci...373..425L}, or 0.318\,count/hour in each bin. We then add the count number of the noise to the event number in each bin so that we obtain a simulated counts map. We then generate the model, composed of another Poisson noise (noise B) based on the measured background rate and a symmetrical 2D Gaussian template (with the normalisation, extension, and positron free) to fit the obtained counts map, and calculate the test statistic (TS) value (TS1) \cite{2017Sci...358..911A,2021ChPhC..45b5002A}. If the TS value exceeds 25, we deem that we identify one source. Next, we set a model composed of noise B plus two symmetrical 2D Gaussian template to fit the obtained counts map again, and calculate the TS value (TS2). If TS2 exceeds TS1 by 25, we deem that we identify two sources. The chain continues by adding another hypothetical source until the increment of TS value is less than 25. 

The formation of mirage sources is related to the magnetic field geometry. After escaping accelerators, electrons spiral around extensive magnetic field lines in the surrounding medium. They are scattered by irregularities in the magnetic field but their trajectories roughly follow the mean direction of the magnetic field over one coherence length $L_{\rm c}$. By definition, the mean field direction changes dramatically in another coherence length. If the field line within certain coherence length roughly orients with our line of sight, a visual projection effect leads to the apparent enhancement of electron column density in that direction, which is not necessarily toward the injection point, creating a mirage source or an offset. This effect is similar to the anisotropic diffusion model described in Ref.\cite{2019PhRvL.123v1103L}.
Note that the turbulent magnetic field is treated as the superposition of many plane waves. Therefore, apart from the main field direction change at the scale of $\sim L_{\rm c}$, there are also smaller bents in the magnetic field within one coherence length. Even if particles do not cross over one coherence length before cooling, they may still form mirage sources or offset due to these small bents.

\subsection{Source numbers}
Due to the stochastic nature of the turbulent magnetic field realizations, the resultant morphology of the pulsar halo and the inferred source structure can vary significantly between simulation runs. A key question is how often one would identify multiple distinct sources around a single accelerator in such mock observations. To address this statistically, we define mirage ratio as $\xi \equiv N-1$ (with $N$ the average source number identified around a particle accelerator), representing the averaged number of extra sources powered by the same particle accelerator.

We perform simulations with 100 randomly generated magnetic field configurations and compute the mirage ratio. We systematically vary parameters including the pulsar distance $d$, the injection spectral index $\alpha$, the ratio of regular to turbulent magnetic field $B_{\rm r}/B_{\rm t}$, and the total magnetic field energy density $B^2$, while keeping the halo luminosity fixed at $L_{\rm ref}$. The resulting trends in the total number of identified sources are presented in \autoref{fig:xi}.

Our results show that the magnetic field strength $B$ has the most dominant influence on the source multiplicity. When $B$ is small, the cooling time of electrons is significantly longer. This allows particles to propagate over larger distances and undergo more isotropic diffusion, effectively washing out small-scale anisotropies. Consequently, the emission morphology tends to be smooth and is predominantly reconstructed as a single, unified source. The influence of other parameters, such as the coherence length $L_{\rm c}$, the injection index $\alpha$, and the ratio $B_{\rm r}/B_{\rm t}$, is present but substantially weaker compared to the decisive role of the magnetic field strength.

\subsection{Source separation and flux ratio}

In this subsection, we discuss the separation distance (defined as the maximal separation distance between any two sources if two or more mirage sources are identified) and the flux ratio ($\chi$, defined as main source/mirage source, where the main source is the source with the smallest distance to the injection point and the mirage source is the farthest one) of the sources. For each chosen set of parameters, we carry out 100 runs and calculate the above two quantities of each run, and show their distribution in the plane of separation value and flux ratio. In all runs, the accelerator's distance is fixed at $d=1\,$kpc. To reveal the difference more clearly, we set here a higher criterion TS$=25$ for the identification of a source. As such, the source number will decrease compared to the case shown in \autoref{fig:xi} with the same parameters (i.e., not all runs can generate a mirage source at $d=1\,$kpc). 

Our results are shown in \autoref{fig:Separation}. In the top panel of the figure, we can see that the mean separation between the main source and the mirage source is positively correlated with $L_{\rm c}$. Given that the mirage source is formed due to the projection effect when particles propagate to another coherence of the magnetic field as discussed before, the separation value is correlated with the coherence length. On the other hand, we do not see a clear dependence of the $N$ on $L_{\rm c}$.

The influence of the magnetic field strength is noticeable on both the separation value and the flux ratio, as shown in the middle panel of \autoref{fig:Separation}. The effect can be attributed to faster cooling of electrons in stronger magnetic fields. Injected electrons may not cross one coherence length before cooling if the magnetic field strength is strong, and therefore only reveal the small bents within one coherence length of the magnetic field, leading to a smaller separation between the main source and the mirage source. Due to the same reason, the electron number consisting of the mirage source is also smaller for a stronger magnetic field, leading to a lower flux ratio. From the middle panel of \autoref{fig:Separation}, we see that the flux of the mirage source may be larger than the main source in a considerable fraction of runs with $B^2=4$ and 9$\,\mu\rm G^2$. For $B^2=36\,\mu\rm G^2$, the flux ratio is greater than unity in most of runs.

The regular-to-turbulent ratio $B_{\rm r}/B_{\rm t}$ exhibits a more complex influence on source separation, as shown in the bottom panel of \autoref{fig:Separation}. In the case of pure turbulent field ($B_{\rm r}/B_{\rm t}=0$), electrons undergo random walk, generating numerous small-scale structures but lacking large-scale coherence, resulting in moderate separation distances. When $B_{\rm r}/B_{\rm t}\approx 1$, an optimal balance is achieved: the regular component provides directional guidance while maintaining sufficient turbulent complexity, leading to well-defined mirage sources with the largest separations. As $B_{\rm r}/B_{\rm t}$ increases to 3, the propagation becomes increasingly dominated by the regular field, suppressing the formation of secondary structures and reducing the separation. This non-monotonic behavior reflects the competition between field-line guidance and turbulent scattering in shaping electron distributions.

\subsection{Offsets}

\begin{figure*}[htbp]
\centering
\includegraphics[width=0.4\linewidth]{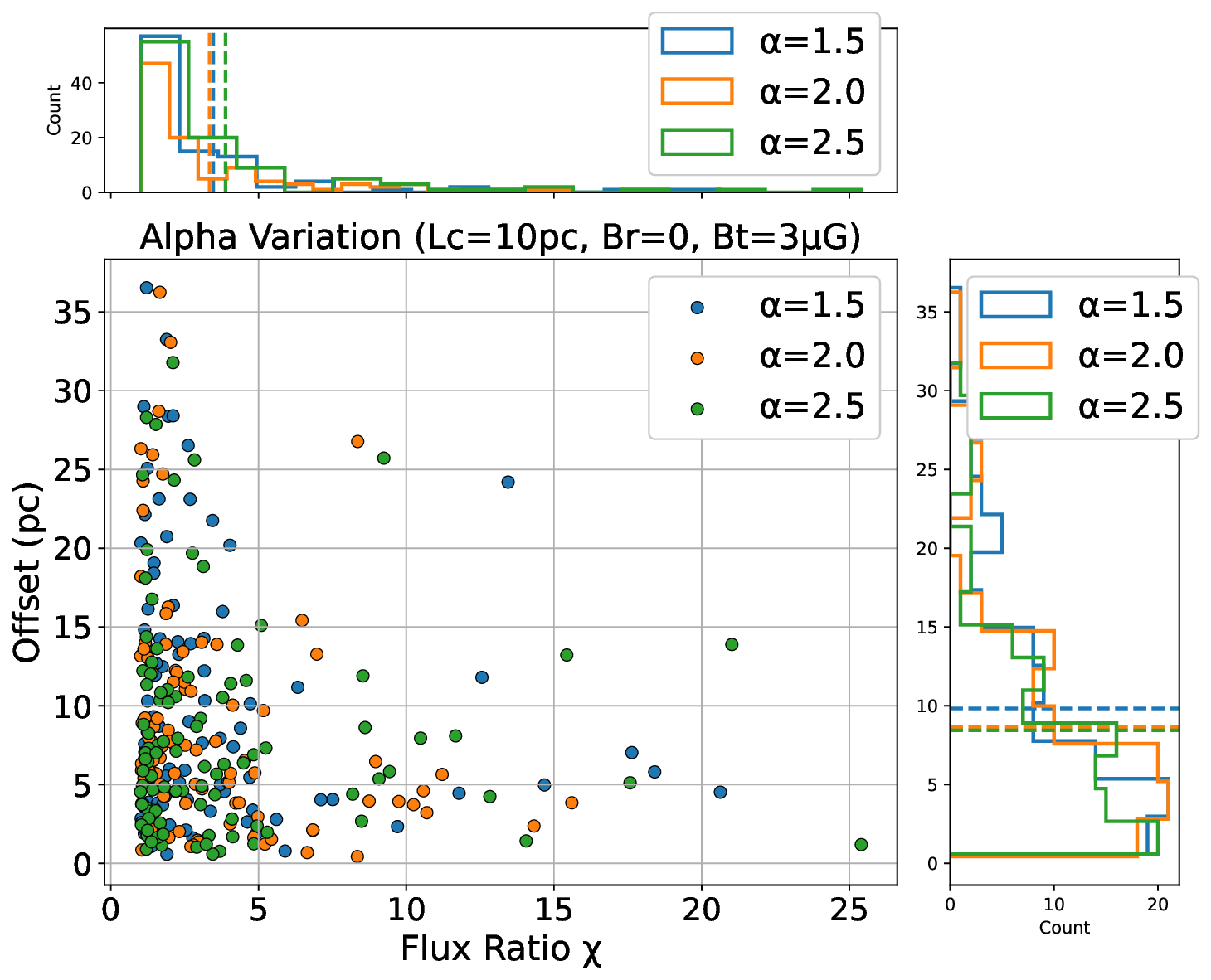}
\includegraphics[width=0.4\linewidth]{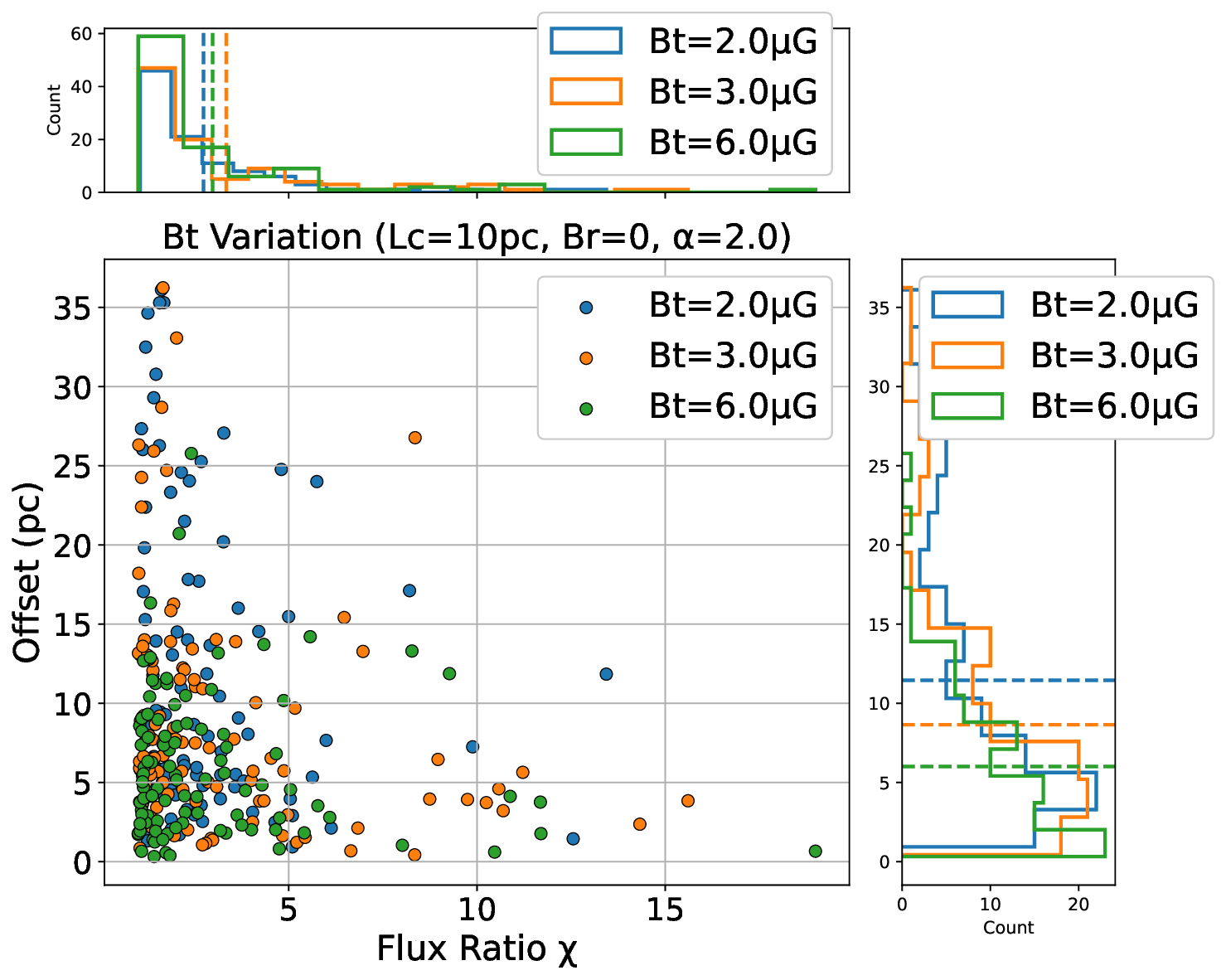}
\includegraphics[width=0.4\linewidth]{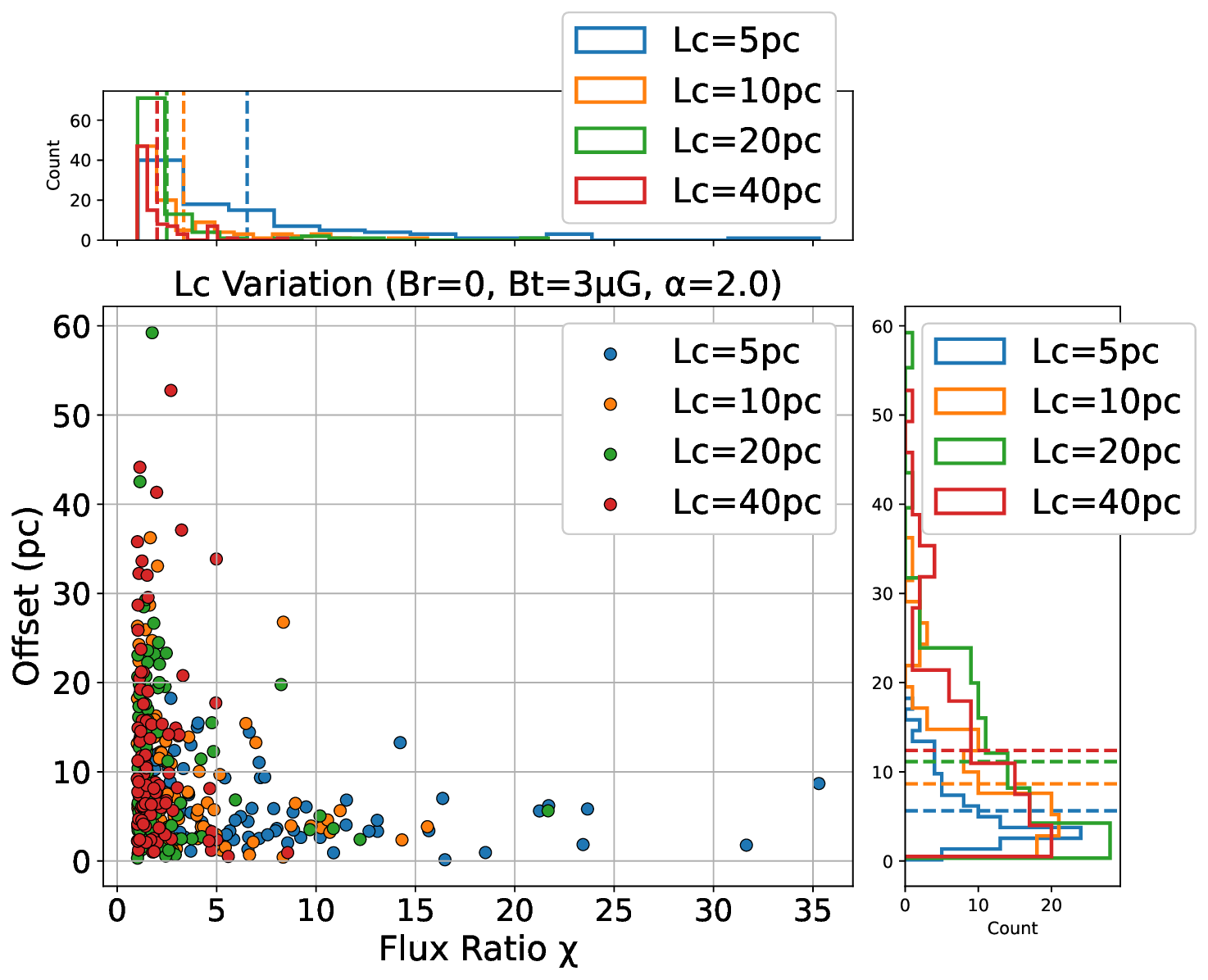} 
\includegraphics[width=0.4\linewidth]{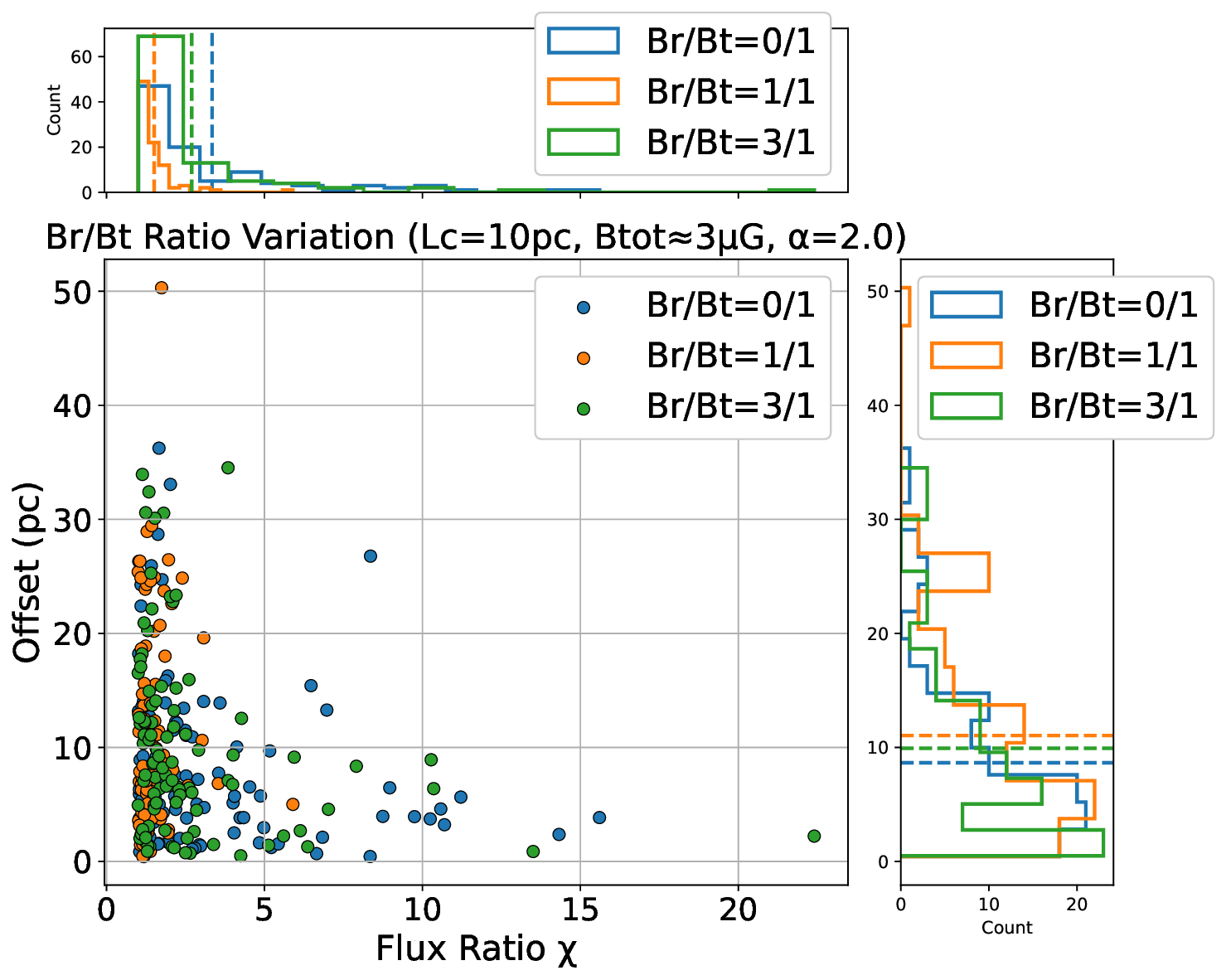} 
\caption{Scatter plot of offsets and total counts of the identified source which is closest to the accelerator for different  $L_c$ (top panels),  $B^2$ (middle panels), and $B_r/B_t$ (bottom panels). The distance is fixed at 6\,kpc (the pixel size 0.1$^\circ$ corresponding to $\approx 10$ pc; the PSF ( 0.3$^\circ$) corresponding to $\approx 30$ pc). In the left columns, the luminosity is $\mathcal{L}=\mathcal{L}_{\rm ref}$, while in the right columns, the luminosity is $\mathcal{L}=10\mathcal{L}_{\rm ref}$. Dashed lines in the histogram mark the average value of the offset. \label{fig:Offset}}
\end{figure*}

Now we investigate the spatial offsets of the identified sources from the accelerator. We locate the source at a relatively large distance, 1\,kpc. Nevertheless, if more than one sources are identified, the offset is defined as the distance between the accelerator and the source closest to it.


The influence of $B_{\rm r}/B_{\rm t}$ on offsets follows a pattern similar to that for source separation. Pure turbulent fields ($B_{\rm r}/B_{\rm t}=0$) produce moderate offsets due to random scattering, while balanced configurations ($B_{\rm r}/B_{\rm t}\approx 1$) yield the largest offsets as electrons maintain directional coherence while exploring multiple magnetic coherence scales. Strong regular fields ($B_{\rm r}/B_{\rm t}=3$) confine electrons closer to the injection point, resulting in smaller offsets.

\section{Summary and Discussion}
Propagation of high-energy cosmic rays in the ISM close to their accelerators plays an important role in shaping the morphology of extended gamma-ray sources.
In the companion paper~\cite{paperI}, we found that the propagation of electrons is highly asymmetric, and that this may lead to the appearance of so-called mirage halos, as well as significant spatial offsets between the positions of the measured gamma-ray sources and that of their parent accelerators. In the present paper, we explored in detail the formation mechanism of these mirage halos and spatial offsets, and we investigated quantitatively the influence of the model parameters on the appearance of these two phenomena. 

We performed test-particle simulations of the propagation of electrons above 100\,TeV in the interstellar magnetic fields around their accelerators, taking radiative cooling into account. With the obtained distributions of electrons, we then calculated their IC radiation and obtained the mock count maps as would be seen by LHAASO, as well as the identifiable sources following the standard likelihood analysis. We found that the mirage sources and the large spatial offsets between identified sources and particle accelerators can be ascribed to the asymmetric propagation of particles in the magnetic field and to projection effects. More specifically, particles propagate preferentially along the direction of their local magnetic field lines, which changes dramatically between two coherence lengths. If the field line within a certain magnetic coherence length is approximately aligned with our line of sight, an apparent enhancement of the electron column density appears in our direction due to a projection effect. A mirage source or a source offset is then created. 

Furthermore, based on the properties of identified sources, the influence of model parameters on the propagation of electrons have been evaluated with three indicators, namely, the source number $N$, the maximum separation distance, and the offset. The maximum separation distance is the maximum distance between any two sources identified in one run. The offset is the distance between the pulsar's position and the centroid of the closest identified source.
Our main findings include:
\begin{itemize}
    \item The $N$ is negatively correlated with $d$, $L_{\rm c}$, and is positively correlated with $B^2$. The other dependences are relatively weak. 
    \item The maximum separation is negatively correlated with $B^2$, and positively correlated with $L_{\rm c}$. It exhibits a peak at intermediate $B_{\rm r}/B_{\rm t}$ ratios, reflecting the competition between field guidance and turbulent scattering. The large maximum separation suggests that in the inter-arm region, and in the Galactic halo (where the coherence length is larger), two sources separated with 100 pc could still be linked with one another and belong to the same halo.
    \item The offset is negatively correlated with $B^2$, and positively correlated with $L_{\rm c}$. Similar to separation, offsets are maximized at balanced $B_{\rm r}/B_{\rm t}$ ratios. Our calculations suggest that in the inter-arm region, and in the Galactic halo (where the coherence length is larger), the offsets between pulsar and the halo centroid could be as large as 50\,pc.  
\end{itemize}
We have also discussed the possible application of this mechanism to explain some LHAASO sources with pulsar associations. For example, 1LHAASO~J0216+4237u, J0212+4254u, and J0206+4302u \cite{2025arXiv251006786C} could have a common origin, and be explained as mirage halos powered by the millisecond pulsar PSR~J0218+4232. The large offsets between a few middle-aged pulsars and LHAASO sources may also be explained by our model naturally.

In our simulations, we do not consider the feedback of relativistic particles on the magnetic field, such as the streaming instability. This simplification is reasonable for energetic electrons ($>100$\,TeV), because the energy density of such high-energy electrons is usually very small compared with that of the interstellar medium, and hence the growth of such instabilities should be very slow. 

We also note that the synthetic turbulent magnetic fields used in our simulations are generated following Ref.~\cite{1999ApJ...520..204G}, assuming an isotropic turbulence with Kolmogorov-type power spectrum. The nature of the turbulence in the interstellar medium may display some differences. In particular, studies have shown that Alfv{\'e}nic turbulence is highly anisotropic at the gyro-scale of cosmic rays and thereby inefficient at scattering cosmic rays, while fast modes are isotropic and should play a dominant role in scattering cosmic rays~\cite{2002PhRvL..89B1102Y}, as supported by the recent analysis of radio polarisation observations \cite{2024ApJ...965...65M}. Therefore, some differences must exist between the cosmic-ray scattering rates for particles propagating in our synthetic turbulence and that for particles propagating in magnetohydrodynamic turbulence. Although our synthetic turbulence may not be accurate on the cosmic-ray gyration scales, it still qualitatively captures the isotropic and filamentary nature of the interstellar turbulence on large scales, which is due to modes with large wavelengths. These modes are actually those which drive the filamentary propagation of cosmic-rays along magnetic field lines and thereby the formation of asymmetric gamma-ray emissions~\cite{2012PhRvL.108z1101G,2013PhRvD..88b3010G}, mirage sources, and large offsets. Therefore, our calculations provide a reasonably good approximation of realistic interstellar turbulence, for the study of the appearance of asymmetric extended gamma-ray emissions around cosmic-ray sources. Furthermore, this method can take into account a wide range of length scales in the turbulence from the gyroradius of relevant electrons up to the turbulence injection scale with reasonable computing times. More realistic simulations, such as particle-in-cell simulations, or particle propagation in 3D cubes from magnetohydrodynamic simulations can, however, only cover a limited range of spatial scales, limiting the ranges of electron energies that can be studied with them.

\begin{acknowledgments}
We thank Chunkai Yu, Hao Zhou and Giovanni Morlino for discussion. This work is supported by 
NSFC under grants No. 12393852, 12350610239, 12393853,  12173018, 12121003, and U2031105, {by the Fundamental Research Funds for the Central Universities under No. 020114380057, and by K. C. Wong Educational Foundation.}
\end{acknowledgments}

\bibliography{apssamp}

\end{document}